\documentclass[apjpt4]{aastex}
\input epsf.sty
\newcommand{\s}{S5~0716+714}

\begin{document}

\title{The Long Term Optical Variability of the BL Lac object 
S5 0716+714: Evidence for a Precessing Jet}

\author{R. Nesci, E. Massaro, C. Rossi, S. Sclavi}
\affil{Physics Dept., University of Roma La Sapienza, \\
Piazzale A. Moro 2, I-00185 Roma, Italy 
}
\email{roberto.nesci@uniroma1.it, enrico.massaro@uniroma1.it}
\author{M. Maesano, F. Montagni}
\affil{Stazione Astronomica di Vallinfreda,  \\ 
Via del Tramonto, I-00020 Vallinfreda, Italy}

\begin{abstract}

We present the historic light curve of the BL Lac object S5 0716+714, 
spanning the time interval from 1953 to 2003, built using Asiago archive 
plates and our recent CCD observations, together with literature data.
The source shows an evident long term variability, over which well known 
short term variations are superposed.
In particular, in the period from 1961 to 1983 the mean brightness
of S5 0716+714 remained significantly fainter than that observed after 1994. 
Assuming a constant variation rate of the mean magnitude
we can estimate a value of about 0.11 magnitude/year. 
The simultaneous occurrence of decreasing ejection velocities of 
superluminal moving components in the jet reported by Bach et al. (2005)
suggests that both phenomena are related to the change of the direction 
of the jet to the line of sight from about 5 to 0.7 degrees
for an approximately constant bulk Lorentz factor of about 12. 
A simple explanation is that of a precessing relativistic jet,
which should presently be close to the smallest orientation angle. 
One can therefore expect in the next ten years a decrease of the mean
brightness of about 1 magnitude.
\end{abstract}

\maketitle

%\newpage

\section{Introduction}

The radio source S5~0716+714 was identified with a bright and highly variable 
BL Lac object, characterized by an intense featureless optical continuum 
\citep{Bie81}.  % (Biermann et al. 1981).
The impossibility to detect any host galaxy suggested that its redshift 
should be greater than 0.3 
\citep{Wag96},  % (Wagner et al. 1996),
however a recent possible detection with XMM-Newton of an X-ray line 
assumed to be Fe XXV (6.7 keV) suggested the lower values of $\sim$0.1 
\citep{Kad04}, % (Kadler et al. 2004), 
but this result has not been confirmed.

Variations on short time scales (a fraction of hour) have been detected 
in several occasions at frequencies ranging from radio to X-rays 
\citep{Gab98,HeiWag96,Gio99,Vil00}.
%(Gabuzda et al. 1998; Heidt \& Wagner 1996; Giommi et al. 1999; 
% Villata et al. 2000).
The main radio and optical properties of \s~ have been recently summarized by 
\citet{Rai03}  % Raiteri et al. (2003) 
which report also an accurate and quite homogeneous photometric data set 
from 1994 to the end of 2001.
On time scales of months and years optical light curves show 
a series of well defined maxima around a magnitude $R\simeq$12.5 
and separated by variable time intervals from $\sim$1 to $\sim$3 years. 

Long term ($\geq$ 10 years) optical light curves are available 
only for a small number of BL Lac objects. 
In some cases these curves generally show a large short-term 
variability superposed onto long-term trends occurring over time 
scales of several decades. An example is that of OQ~530 whose
optical brightness is fading since about one century 
\citep{Mil78,Mas04}.  % (Miller 1978; Massaro et al. 2004).
The origin of the fast variations is generally explained by the
relativistic boosting of perturbations moving down a jet pointing 
close to the line of sight. The relevant quantity is the beaming
factor $\delta = 1/\Gamma(1-\beta~cos\theta)$, where $\Gamma$
and $\beta$ are the Lorentz factor and the velocity (in units of the
speed of light) of the perturbations' bulk motion and $\theta$ is
the angle between the jet and the line of sight.

On the other hand the nature of secular variations is unclear. 
A suggestive possibility is that they can be reasonably associated 
with changes in the structure and/or direction of the inner jet. 
It is difficult, however, to obtain a clear evidence of such changes
because it requires long and accurate multifrequency campaigns on
a sample of several sources, whereas large data sets are available
only for a few BL Lacs.

We focused our attention on the long-term behavior of \s~ for
which a detailed study of proper motions in the jet, based on
a quite dense VLBI image database, has been recently performed by 
\citet{Bac05}.  % Bach et al. (2005).
Optical photometric data since 1994 indicate the existence of 
a slow increasing of the mean luminosity. 
To verify this indication we searched for historic images in 
astronomical archives using the Wide-Field Plate Database at CDS
and found a number of Schimdt plates, covering the period from 
1961 to 1985 in the Asiago Observatory archive. 
In this paper we report the results of the photometric measurements
of these images and present the first optical historic light curve
of \s. 
In addition we report the results of our recent CCD observations
not included in previous papers.
As we will show, these data confirm the existence of secular luminosity 
changes in BL Lac objects and rise the opportunity to preserve and organize 
archives of historic images to investigate variable phenomena over time 
scales longer than a single generation of astronomers.
We also compare our findings with the superluminal motion of jet
components derived from 
\citet{Bac05} % Bach et al. (2005) 
and show that they can be combined together in a consistent scenario of 
a slowly precessing jet. 

\section{Photometric Observations and Data Analysis}

\subsection{The recent activity: 1995--2003} \label{recent}

The optical brightness of \s~ has been monitored since the beginning of
nineties by a wide international collaboration: a large data set of
multiband photometric measurements up to the end of 2001 is given 
by \citet{Rai03}. % Raiteri et al. (2003). 
We continued to observe \s~ and report here new data taken in the 
years 2002 and 2003.    
Our observations were performed with two small aperture reflector 
telescopes equipped with CCD cameras. 
The used telescopes were the 50 cm telescope of the Astronomical Station 
of Vallinfreda (Roma) with an Apogee camera mounting a back-illuminated 
Marconi 47-10 chip and a 31 cm f/4.5 newtonian reflector located 
near Greve in Chianti (Firenze) equipped with a DTA HiRes II CCD camera
mounting a back-illuminated SITe SIA502A chip.
Standard $B, V $(Johnson) and $R, I$ (Cousins) filters were used. 
Exposure times depended upon the the brighness of the source and varied
between 3 and 5 minutes. Aperture photometry was performed using a 5 
arcsec radius with IRAF/APPHOT and calibrated magnitudes obtained using 
the photometric sequence defined in the field by 
\citet{Vil98}. %  Villata et al. (1998). 
Our new data are therefore on the same scale of those published in the 
database by
\citet{Rai03}. % Raiteri et al. (2003).
Errors $\leq$0.03 on the magnitude of the source were always achieved.

Figure \ref{LCR0716n} summarizes the recent brightness history of \s: 
$R$ magnitude was generally in the range 14.5 - 13 with the exception 
of a small number of short flares in which the source was brighter 
up to $R$=12.3. 
The time separation between the flares was around 1200 days and seems
to decrease to about 300 days after 2001. A remarkable fact is that
\s~ was observed fainter than $R$=14.5 only in a very few occasions,
and in the last three years it was never found fainter than 14.   
This behavior suggests that the source has a long term brightening
detectable only when well sampled light curves are available.
Note also that the peak brightness reached in the outbursts seems
to follow a similar trend.

A simple linear fit applied to these data allows to estimate a mean 
brigthening of about 0.09 mag/year, which corresponds to changes 
of the mean magnitude larger than 1 mag, when considered over time 
intervals longer than ten years.
It is important to verify if this trend will continue in the next years
because \s~ would be the brighter BL Lac object ever observed.
 
\subsection{Historic photographic data: 1961--1985} \label{historic}

A possibility to confirm this long term behavior is to search
for photographic images of \s~ taken before 1990.
We found 53 useful plates in the Asiago Observatory archive, containing 
our source: 46 of them were obtained with the 50/40/120 cm Schmidt 
telescope and 8 with the 90/67/245 cm one. 
These plates were taken as part of the Supernova sky patrol of
the Asiago Observatory and were never used before for the study of \s.
The covered time interval is from February 1961 to January 1985, but with  
a highly uneven sampling. Four images were taken in 1961, four in 1965-1966, 
the other 45 are distributed in the interval between 1973 and 1985.

The magnitude evaluation was not simple because of the non homogeneity
of filter/emulsion combinations, characterized by different effective 
wavelengths: 103aO+GG13 (closely matching the Johnson's B filter), 
103aO (3600-5000 \AA), TriX (3600-6700 \AA) and Panchro Royal (3800-6400 \AA). 
Plates were digitized at the Asiago Observatory with an EPSON 1680 
Plus scanner as part of a national program of digitization of 
astronomical plate archives 
\citep{Bar03}.   % (Barbieri et al. 2003).
A sampling step of 16 micron (1600 dpi) was used, in grayscale/transparency 
mode and 16 bit resolution.
Plate scanning included also the unexposed borders to measure the 
plate fog level $F$. Due to the presence of a residual scattered light 
in the scanner, we found that the instrumental zero value $Z$ for each plate 
was better evaluated using the central pixels of the most overexposed star.
The transformation of the recorded plate transparency $T$ of each pixel 
into a relative intensity $I$ was obtained applying the simple relation 
$I=(F-Z)/(T-Z)$.

We found the source in the Asiago plates significantly fainter than in the
recent CCD images and consequently we had the problem to extend the  
photometric sequence of 
\citet{Vil98},   % Villata et al. (1998),
limited to $B$=14.7, to fainter stars to obtain reliable estimates of the 
brightness of our target.
We then selected 23 stars ranging from $B$=11.5 to 17.7 and used 
12 CCD frames of the field obtained with the Vallinfreda telescope 
to perform a calibration of this new sequence.

We checked that the selected stars were not variable and that their 
$B-V$ color indices were similar to the typical one of our source 
\citep[$B-V \simeq$ 0.45,][]{Rai03},
% ($B-V \simeq$0.45, Raiteri et al. 2003), 
to minimize any possible color effect in deriving its $B$ magnitude.
Four stars were found to be variable, two of them having
quite red colours ($B-V \ge 1$), and were excluded from the 
reference set, leaving us with a final list of 19 stars. 
A finding chart for these stars is given in Fig. \ref{cartina}, where the
reference stars are marked with small letters and \s~ is marked with S5.
Table \ref{std0716} gives the star flag (column 1), 
RA and DEC (J2000) derived from the Palomar DSS2 (columns 2 and 3), 
the adopted B magnitude (column 4) with its uncertainty (column 5),
the adopted V magnitude (column 6) with its uncertainty (column 7)
and the correspondence to the \citet{Vil98} sequence, if existent (column 8).

\begin{figure}
\epsscale{1.0} 
\plotone{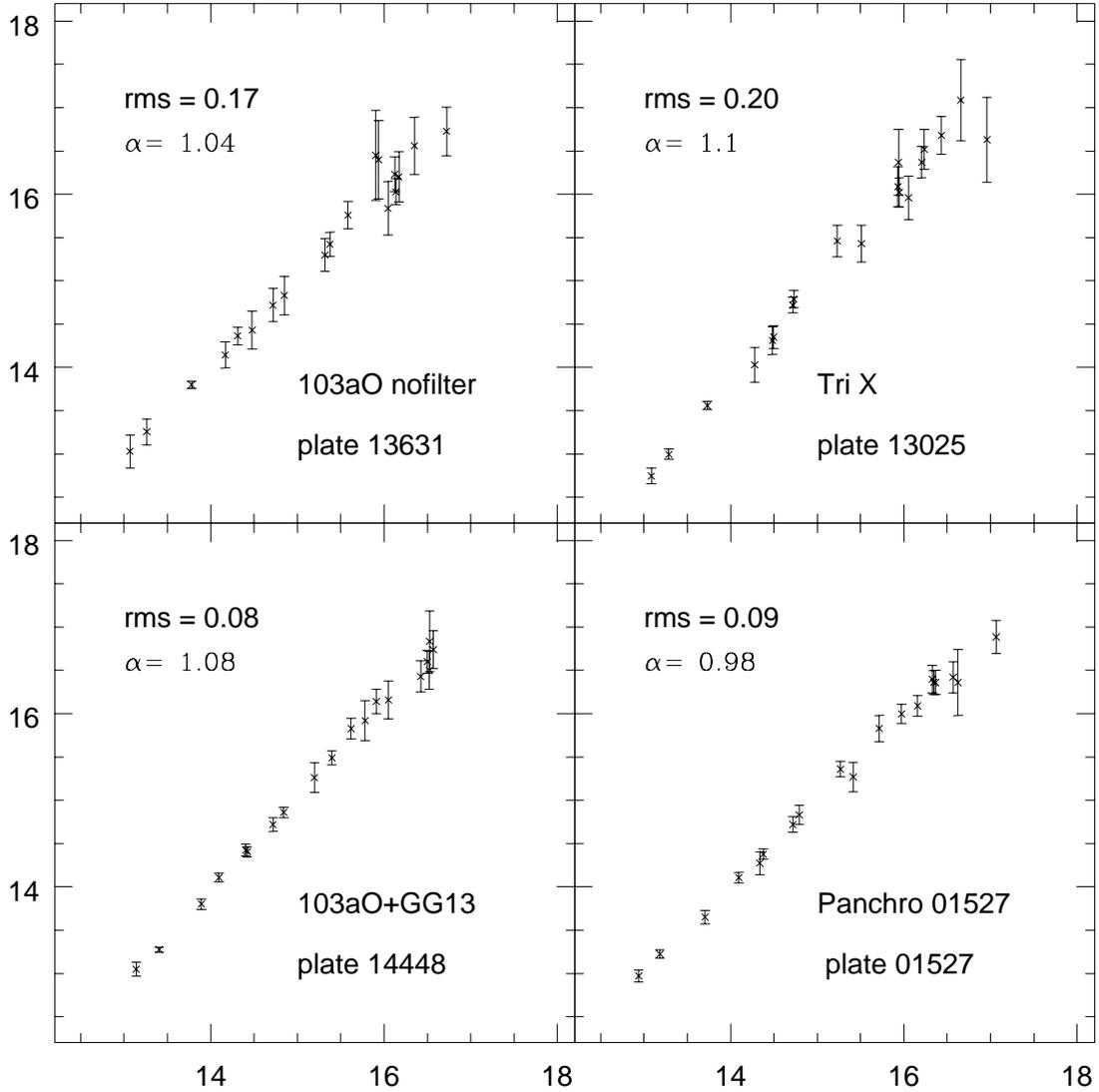}
\caption{Consistency check for different plate/emulsion combinations.
Abscissae are $B$ magnitudes of a well exposed plate; ordinates are
the average magnitudes of the other plates. Error bars are the rms deviations
of each star for the plate subset.
\label{40tutted}
}
\end{figure}

Instrumental magnitudes for each plate were obtained using IRAF/APPHOT 
tasks with a fixed aperture of 2.5 pixels for the 40cm Schimdt plates and 
4.0 pixels for the 67cm plates, corresponding to the FWHM of the average 
stellar profiles.
For each group of plate/emulsion combination we constructed the
scatter plot between the magnitudes obtained from a well exposed plate 
and the mean value of the magnitudes on the other plates. 
The results for the 40cm Schmidt plates are shown in Fig. \ref{40tutted}. 
We remark that all the data are well fitted by straight lines
indicating a good consistency of the data set.
For each star the scatter around the average magnitude is shown.
The same check for the 67cm Schmidt plates gives even better results.

\begin{figure}
\epsscale{1.0} 
\includegraphics{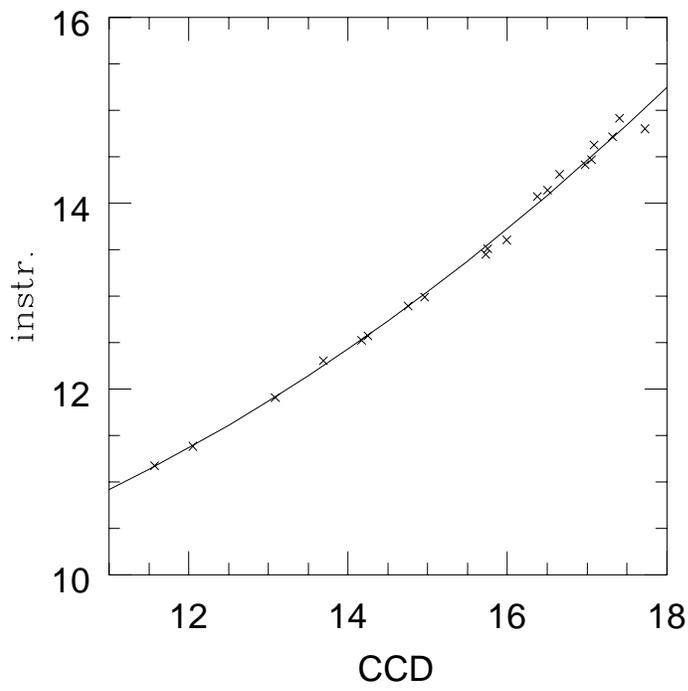}
\caption{Calibration curve for the plate S40-01464. 
Abscissa is CCD $B$ magnitude for the reference stars;
ordinate is the instrumental photographic value. 
\label{calccd}
}
\end{figure}

A calibration curve for each plate was then obtained using the CCD 
values of the reference stars. 
These curves were remarkably tight and showed a slight departure from 
a simple linear relation, so we fitted them with a 2nd order polynomial 
law.
The scatter of the points around the fitting curve depend only 
on the plate quality, without systematic dependence from the filter/emulsion
combination: a typical calibration curve is shown in Fig. \ref{calccd}.
The $B$ magnitude of \s~ was then derived from the instrumental one
using the fitted calibration curve.
Typical rms errors are 0.1 mag for the 67cm Schmidt, while for the 40cm 
Schmidt they range from 0.15 at $B$=14 to 0.30 at $B$=17.5. 
In several nights, two consecutive plates were taken by the 40cm 
Schmidt with different emulsions: no systematic differences in the 
derived magnitudes were found, at our precision level. 
We averaged therefore data referring to the same night to build the 
light curve.
The list of $B$ magnitudes and relative uncertainties of \s~ from the 
Asiago Observatory archive is given in Table \ref{histBmag}.
 
Two additional points were obtained from the POSS 1 (year 1953) digitized 
blue plate and from the Quick V (year 1983) plate, both available on-line.
For these magnitude estimates we used the same reference stars and data 
reduction procedure used for the Asiago plates, save that $V$ magnitudes 
were used to calibrate the photometric sequence on the Quick V plate 
and the $V$ magnitude of \s~ was converted to $B$ assuming $B-V$=0.45 
\citep{Rai03}.  % (Raiteri et al. 2003).
Such average value is confirmed also from our photometric data after 
year 2001.

\section{The long term behavior of S5 0716+714} 
The most complete history of the optical brightness of \s~ in the $B$ band
published to now, spanning a time interval from 1953 to 2003, is shown 
in Fig. \ref{bt0716h}. 
Data up to 1985 are from the Asiago plates (plus two Palomar plates) 
described in Section \ref{historic}, measurements from 1989 up to the end 
2001 have been extracted from the data base by Raiteri et al. (2003), whereas 
the measurements from 2001 to 2003 are new CCD data obtained by our group 
(see Section \ref{recent}). These data are available in electronic form at 
the CDS, in the same format of the 
\citet{Rai03}  % Raiteri et al. (2003) 
database.

The most remarkable feature of the old portion (years 1961-1985) of the 
light curve is that its average brightness is significantly fainter than
that recorded in the years 1994-2003. 
This is an important confirmation that the long term trend detected in 
the last ten years started in the eighties.
Short term variability appears to be present also in the years 
1961-1985 with an amplitude comparable to that observed after 1994.
Although the rather poor sampling, the data scatter indicates that
the source was not quiet, or switched off, in the past, but remained
on the average in a less bright state.
Note also that the typical magnitude variation rate ($\sim$0.11 mag/year, 
dotted line) is remarkably close to that estimated in Section \ref{recent} 
from the $R$ light curve after 1994, indicating that the evolution of the 
source brightness continued over such a long time interval.

\section{Discussion}
The results on the historic light curve of \s~ reported for the first
time in this paper confirm that also this bright BL Lac object presents
long term trends with typical time scale of about 10 to 15 years.
A similar behavior is shown by some other well studied BL Lac sources, 
like ON~231 \citep{Mas01}
whose mean luminosity was brightening for about 25 years until the great 
outburst occurred in April 1998 \citep{Mas99} 
to enter afterwards in a slowly fading phase. 
The historic light curve of OQ~530 \citep{Nes97,Mas04} 
can be traced back to the first decades of the past century 
\citep{Mil78} and shows a slowly decreasing brightness.
More complex are the luminosity variations of OJ~287 characterized
by recurring bursts with a period of 12 years \citep{Sil88}. 
Superposed to the bursts, a long term decreasing trend, possibly started
around 1970--1975 and lasting at least since 25 years, is well apparent. 
After year 2000 this trend turned in a faster brightening 
\citep{MasMan04}.

The origin of the long term trends observed in some bright BL Lac objects 
is not fully understood. 
This phenomenon is clearly different from fast and strong bursts 
associated with relativistic boosting of perturbations moving down a jet 
closely aligned to line of sight 
\citep{BlaRee78}. 
Long term changes involve spatial dimensions of several light years 
and therefore they are related to slow changes of the BL Lac nucleus.
As recently pointed out by \citet{MasMan04},
important informations on the nature of these changes can be obtained by a 
comparison of long term optical behavior with the evolution of
some structural parameters derived from VLBI imaging. 
An interesting possibility would be to search for changes of the
jet direction as expected from a precession motion.
In the case of OJ 287, for instance, \citet{TatKin04}
found from the images of the geodetic database 
(RRFID, http://rorf.usno.navy.mil/rrfid.shtml) 
that its jet rotated of about 30 degrees in 8 years.

\begin{figure}
\epsscale{1.0}
\includegraphics[scale=0.70]{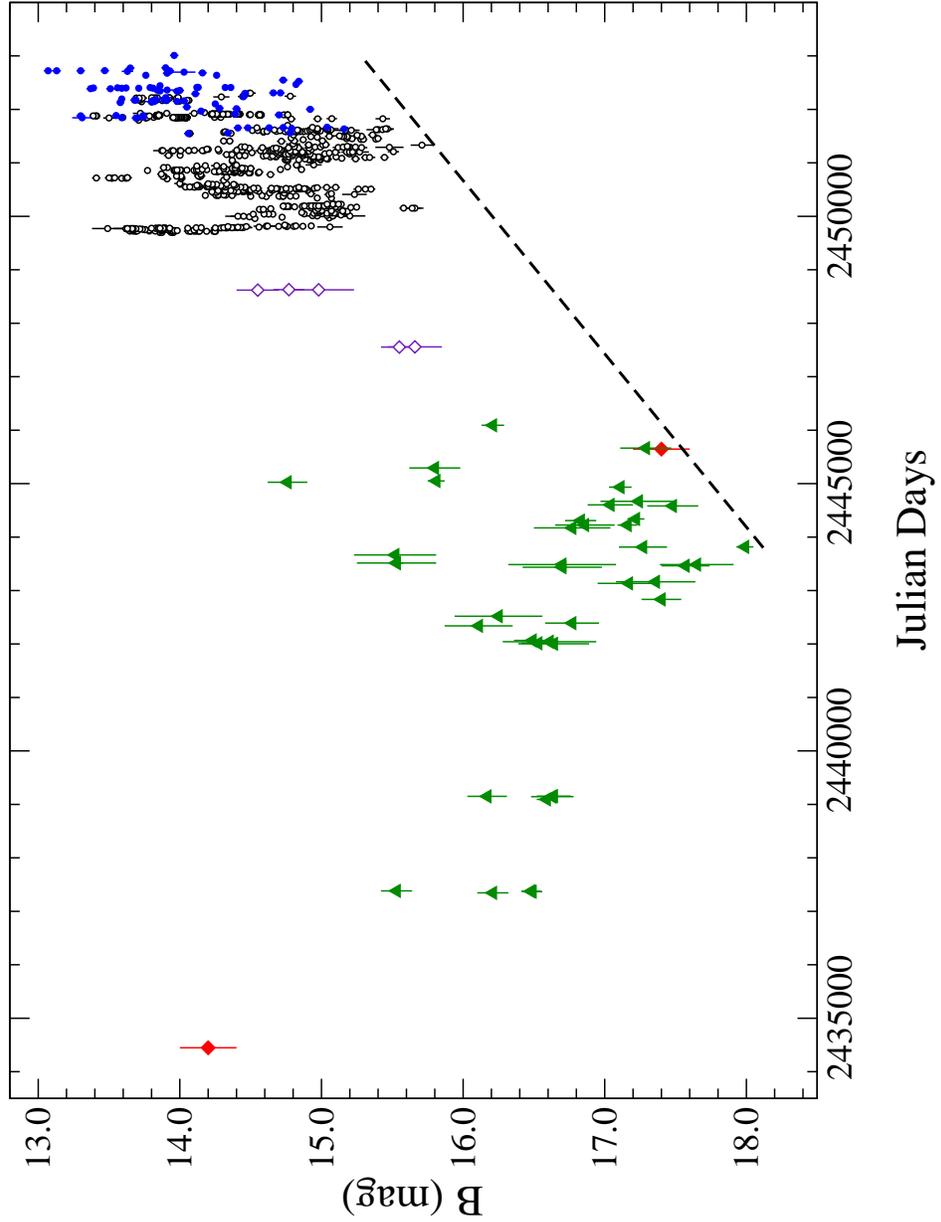}
\caption{The historic light curve in the $B$ band of S5 0716+714 
from February 1953 to March 2003. Open simbols are literature data
\citep{Rai03}:
circles are data from the CCD monitoring program and diamonds from the 
Torino plates archive; filled simbols correspond to new photometric data 
(circles) and to Asiago archive data (triangles). 
Two additional point from POSS-1 and Quick-V plates are also plotted
(filled diamonds).
The dotted line represents a possible long term trend from the 
minimum observed in the years 1975-1980 to the present high state.  
\label{bt0716h}
}
\end{figure}

\begin{figure}
\epsscale{1.0}
\includegraphics[scale=0.70]{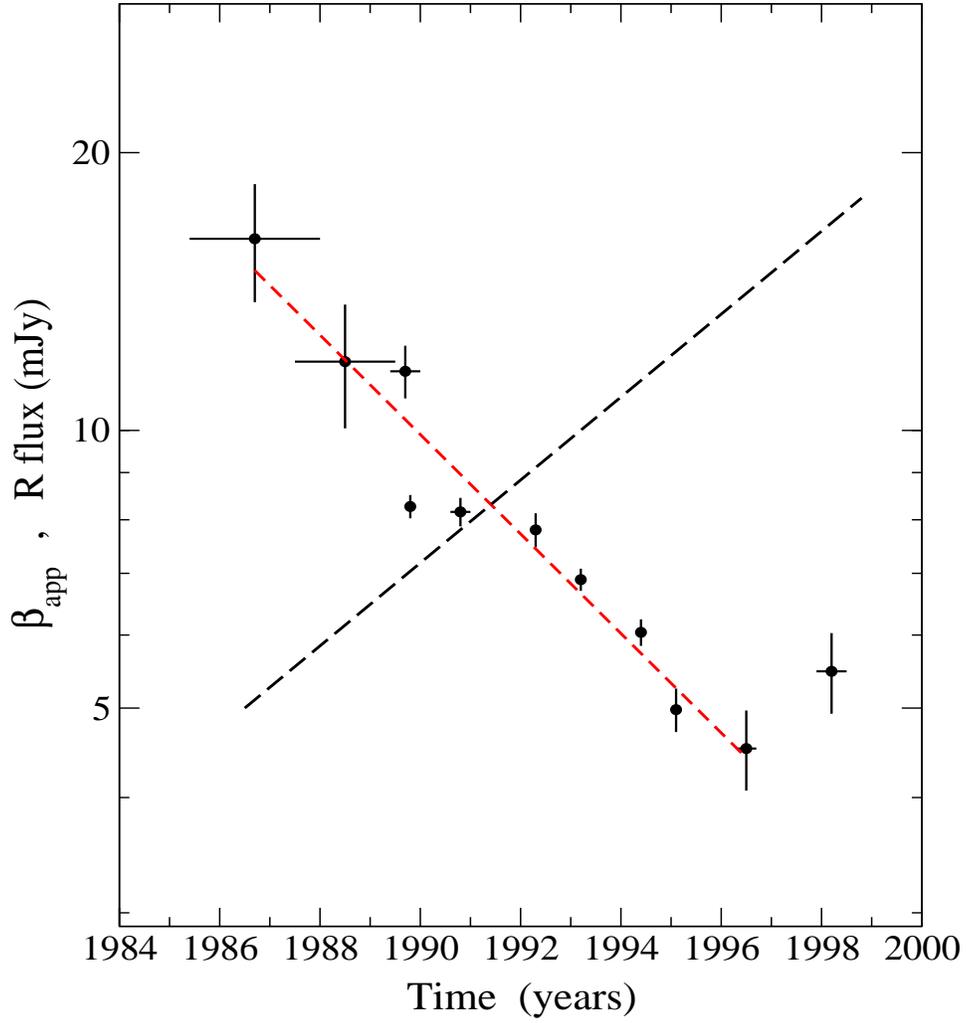} 
\caption{The ejection velocity $\beta_{app}$ of VLBI components in the jet 
of \s~ derived by \citet{Bac05} plotted as a function of time. 
The short dashed line is a best fit of the data with the exclusion 
of the last point.
The long dashed line is the average trend of the optical flux in the
$R$ band assuming the historic trend of Fig. \ref{bt0716h}.
\label{betastor}
}
\end{figure}

\citet{Bac05} 
have recently analyzed the historic data set of VLBI images of \s~ 
at several frequencies from 1992 to 2001 and studied the proper motion 
of several components in the jet. 
In particular, they evaluated the ejection dates of the components 
on the basis of their superluminal apparent velocities 
$\beta_{app}=(\beta~sin\theta)/(1-\beta~cos\theta)$ 
estimated for an assumed resdshift of 0.3. 
In Fig. \ref{betastor} we report the measured values of $\beta_{app}$ 
as a function of the estimated ejection epochs of the various 
components, which show an evident decreasing trend from 16.1 in 1986 
to 4.5 in 1996.
In the same figure we also plotted the evolution of the mean optical 
flux in the $R$ band derived from the long term trend of the historic
light curve of Fig. \ref{bt0716h}, corresponding to a brightening of about 
1.4 magnitudes in the 12 years interval.
 
\begin{figure}
\epsscale{1.0}
\includegraphics[scale=0.70]{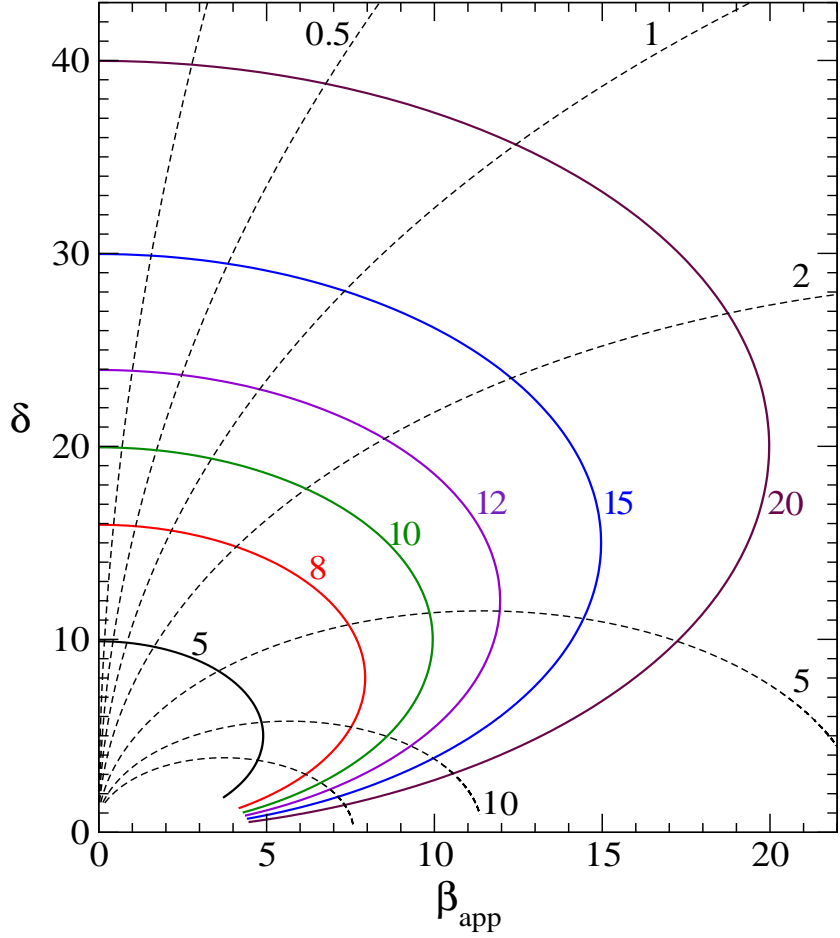}
\caption{The beaming factor $\delta$ plotted vs. the apparent velocity 
$\beta_{app}$ along the jet. Solid lines correspond to constant Lorentz 
factors $\Gamma$ whose values are indicated in the figure; 
dashed lines correspond to constant orientation angles $\theta$ of the jet 
direction to the line of sight varying from 0.2 to 15 degrees.
\label{betadelta}}
\end{figure}

The possibility to observe these apparently opposite behaviors over time
scales of 10--20 yr or longer has been recently pointed out by 
\citet{MasMan04}  
as a useful test to detect a slowly precessing jet in a blazar source. 
It is a direct consequence of a regular change of the angle $\theta$ 
between the jet direction and the line of sight.
The $\delta~ {\rm vs} ~\beta_{app}$ plot of Fig. \ref{betadelta} shows that 
a decreasing ejection velocity $\beta_{app}$ simultaneous to an increase of 
the beaming factor $\delta$ can be realized when the jet is approaching 
to the line of sight, i.e. when one is moving toward left on the upper 
branches of the constant $\Gamma$ curves. In particular, as already stated by 
\citet{Bac05}, % Bach et al.(2005),
the observed decrease of $\beta_{app}$ would imply a Lorentz factor of
about 12--15 and $\theta$ varying from about $5^{\circ}$ to about 
$0^{\circ}.7$. In this conditions we expect that $\delta$ increases
by a factor of about 1.5, from 15 to 22 (see Fig. \ref{betadelta}), 
with a corresponding change of the flux density (proportional to 
$\delta^{3+\alpha}$, where $\alpha$ is the spectral index) around 4, 
well comparable with that derived from the long term trend.

\begin{figure}
\epsscale{1.0} 
\includegraphics[scale=0.70]{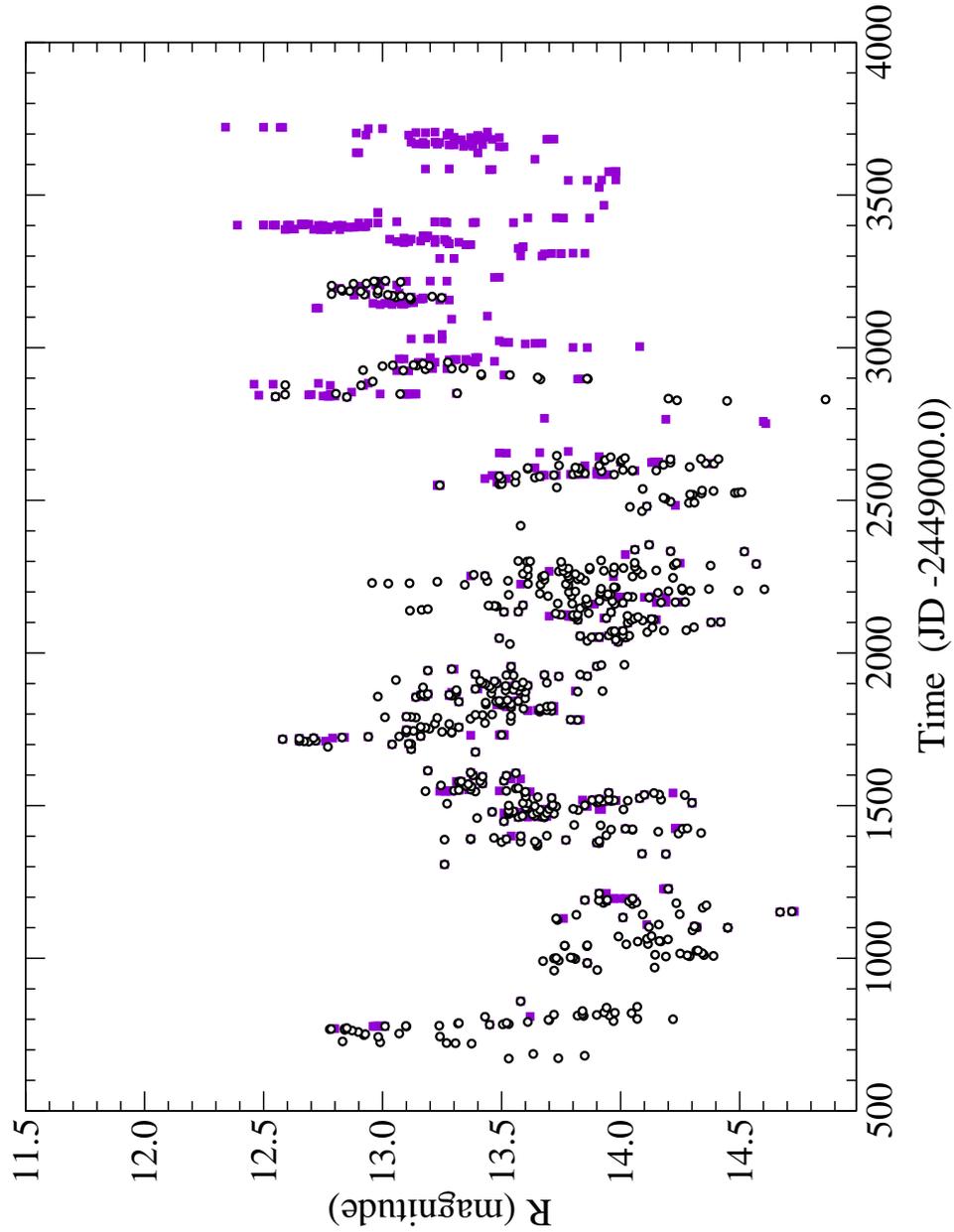}
\caption{The light curve in the $R$ (Cousins) band of S5 0716+714 
from February 1995 to March 2003. 
Open circles are data from 
\citet{Rai03},   % Raiteri et al. (2003),
taking only one measurement per day, filled squares are our new measurements.  
Typical observational uncertainties are of a few hundreds of 
magnitude and error bars are omitted for clarity. 
\label{LCR0716n}
}
\end{figure}

One can also expect that the ejection of radio components in the jet
would correspond to the occurrence of an optical flare. 
A similar effect was found for other blazars by \citet{Sav02} 
who found a correlation between an increase in the core radio 
flux and the appearence of a new born component in the nuclear 
environment.
In the case of \s~ a dense enough optical light curve covers only 
the time interval after the end of 1994, and therefore it is possible
to take into consideration only the last three ejection dates derived by 
\citet{Bac05}. 
The component originated at the beginning of 1995
is indeed coincident with the first outburst in the optical light curve
(about at day 750 in Fig. \ref{LCR0716n}). The ejection date of the most 
recent component has been estimated about five months after the optical burst 
occurred in September 1997 (day 1717 in Fig. \ref{LCR0716n}): considering 
the relatively large uncertainty on the date ($\sim$4 months) one 
cannot exclude that they are related. 
For the radio knot originated around 1996.5 we note that it lies 
within a gap of about three months in the optical light curve: 
the optical flux remained for several months lower than $\sim$10 
mJy before the gap and was higher after, therefore we cannot exclude 
that a rapid flare occurred. 
On this small statistical basis it is not possible to safely establish
a relation between the ejection of superluminal knots in the jet 
and optical flares.
It will be interesting to verify whether new radio knots have been 
originated in correspondence of the recently observed flares in 
November 2000, April 2002 and March 2003.
 
The scenario of a precessing relativistic jet has been proposed to explain 
the evolution of the jets' structure observed in other BL Lac sources 
(e.g. OJ~287, \citeauthor{Abr00} \citeyear{Abr00}; 
BL~Lac, \citeauthor{Sti03} \citeyear{Sti03}) and blazars
(e.g. 3C~120, \citeauthor{CapAbr04a} \citeyear{CapAbr04a}; 
3C~345, \citeauthor{CapAbr04b} \citeyear{CapAbr04b}). 
A precession motion with a period ranging from a few to tens of years can 
be produced in a binary system of massive black holes 
\citep{Kat97}.
Orbital motion of black holes could also be responsible of periodic
outbursts in the optical light curve: indication of possible periodicities
has been reported for a number of BL Lac objects, the first case
being that of OJ~287 \citep{Sil88}. 
 
\citet{Rai03} reported for \s~ a possible recurrence time of 
$\sim$3.3 yr  in the optical light curve, which however is not
apparent at radio frequencies. The most recent flares (Fig. \ref{LCR0716n}) 
have a time separation of about 1 yr, suggesting that the above finding
could not be confirmed. Because we are mainly interested in the 
long-term trend we did not performed any search for periodic
effects: to reduce windowing effects such a study requires a well 
sampled light curve and could better be carried out when all the data
obtained in on-going observational campaigns 
\citep{Ost04} will be available.

In conclusion, historic optical data and VLBI monitoring of \s~ provide
a new significant evidence for the existence of a precessing jet. 
Furterhmore, we found that the estimates of a small viewing angle 
($\theta \leq 2^{\circ}$) and of a beaming factor of $\sim$ 15--20 given by 
\citet{Bac05} are well consistent with the mean increase of optical flux.

It will be interesting to verify in the next years how the mean luminosity 
of \s~ will continue to evolve.
According to the previous estimate of the last $\theta$ value, the jet is 
rather close to a perfect alignment and consequently it is likely 
that $\theta$ will soon start to increase. 
We expect, therefore, the beginning of a dimming phase of the
mean luminosity whereas the ejection velocities of superluminal components 
into the jet will start to increase.

This result confirms that multifrequency monitoring of this BL Lac objects,
and generally of blazars, is important to unravel the physical mechanism
working in these sources. 
A difficulty is due to the occurrence of fast and large outbursts which 
make quite hard the study of time series.
The only possibility to overcome this problem is to extend the observations
over time intervals of several decades to filter the effects of rapid 
variations. 
It is also important to extend the sample of studied objects
to other types of blazars to verify if and how such long term behaviors
are generally detectable or not. 

%\bigskip

%\bigskip
%newpage

\begin{acknowledgements}

We are grateful to F. Mantovani for useful discussions.
Part of this work was performed with the financial support of the Italian 
MIUR (Ministero dell' Istruzione, Universit\`a e Ricerca) under the grants 
Cofin 2001/028773, 2002/024413 and 2003/027534. This research made use of 
the CDS (Strasbourg) database facility and of the Digitized Sky Survey plates.
\end{acknowledgements}

%\newpage

\clearpage

% Tabella 1
%\vspace{-3.0cm}
\begin{deluxetable}{cccccccc}
\tabletypesize{\footnotesize}
\tablecolumns{8}
\tablewidth{0pc}
\tablecaption{Reference stars in the field of S5 0716+714 \label{std0716}}
\tablehead{
\colhead{Star} & \colhead{RA(2000)}& \colhead{DEC(2000)} & \colhead{$B$} &
\colhead{$\sigma$(B)}  & \colhead{$V$} & \colhead {$\sigma$(V)} & \colhead{Villata et
al. (1998)}\\
\colhead{}     & \colhead{}     & \colhead{}          & \colhead{}        &
\colhead{}     & \colhead{}     & \colhead{}          & \colhead{identification}\\
}
\startdata
\hline\\
  a  & 07 21 28.86 & +71 22 56.3 &  17.73  &  0.17  & 16.86 & 0.05 &         \\
  b  & 07 22 17.95 & +71 23 34.7 &  13.69  &  0.03  & 13.23 & 0.02 &   4     \\
  c  & 07 22 36.39 & +71 22 48.7 &  11.57  &  0.04  & 11.07 & 0.02 &   1     \\
  d  & 07 21 23.80 & +71 21 28.1 &  16.50  &  0.06  & 15.96 & 0.04 &         \\
  e  & 07 22 12.56 & +71 21 15.0 &  14.25  &  0.02  & 13.65 & 0.02 &   6     \\
%%f  & 07 21 53.32 & +71 20 36.1 &  14.27  &  0.40  & 13.83 & 0.40 &  S5 0716+714         
\\
  g  & 07 21 33.27 & +71 19 19.9 &  12.05  &  0.01  & 11.53 & 0.02 &   2     \\
  h  & 07 21 41.14 & +71 19 11.7 &  15.75  &  0.05  & 15.06 & 0.03 &         \\
  i  & 07 21 54.36 & +71 19 20.7 &  14.17  &  0.01  & 13.58 & 0.01 &   5     \\
  j  & 07 21 20.71 & +71 18 47.3 &  16.38  &  0.08  & 15.78 & 0.02 &         \\
%%k  & 07 22 21.70 & +71 19 41.2 &  17.53  &  0.11  & 16.36 & 0.04 &         \\
  l  & 07 21 52.22 & +71 18 16.8 &  13.09  &  0.02  & 12.48 & 0.01 &   3     \\
  m  & 07 22 10.49 & +71 17 49.2 &  16.66  &  0.05  & 16.06 & 0.03 &         \\
  n  & 07 22 31.53 & +71 17 48.8 &  16.98  &  0.08  & 16.34 & 0.04 &         \\
  o  & 07 22 28.17 & +71 17 37.0 &  14.75  &  0.02  & 14.18 & 0.02 &   8     \\
%%p  & 07 21 32.67 & +71 16 38.1 &  17.05  &  0.23  & 16.50 & 0.03 &         \\
%%q  & 07 22 06.75 & +71 16 51.0 &  17.34  &  0.11  & 16.20 & 0.03 &         \\
%%r  & 07 22 13.97 & +71 16 43.0 &  16.76  &  0.07  & 16.74 & 0.04 &         \\
  s  & 07 21 37.55 & +71 15 50.6 &  17.09  &  0.12  & 16.52 & 0.01 &         \\
  t  & 07 21 41.75 & +71 15 00.3 &  15.99  &  0.06  & 15.24 & 0.01 &         \\
  u  & 07 21 53.47 & +71 14 45.6 &  15.73  &  0.04  & 15.12 & 0.03 &         \\
  v  & 07 22 40.15 & +71 15 13.3 &  14.96  &  0.04  & 14.34 & 0.04 &         \\
  x  & 07 23 26.00 & +71 15 49.8 &  17.32  &  0.11  & 16.72 & 0.04 &         \\
  z  & 07 22 35.16 & +71 14 39.9 &  17.41  &  0.13  & 16.84 & 0.08 &         \\
\enddata
\end{deluxetable}

% Tabella 2
\begin{deluxetable}{lcccclcc}
\tabletypesize{\footnotesize}
\tablewidth{0pc}
\tablecolumns{8}
\tablecaption{Historic $B$ magnitudes of \s~ from the Asiago archive 
\label{histBmag}}
\tablehead{\colhead{Julian Day} &\colhead{$B$} &\colhead{$\sigma_B$} & & &
\colhead{Julian Day} &\colhead{$B$} &\colhead{$\sigma_B$} \\}
\startdata
\hline\\
 24434447.08 & 14.20 &  0.20 & & & 24443514.43 & 15.53 &  0.28 \\
 24437345.55 & 16.21 &  0.11 & & & 24443514.49 & 15.53 &  0.05 \\
 24437367.47 & 16.49 &  0.07 & & & 24443666.46 & 15.52 &  0.29 \\
 24437375.33 & 16.48 &  0.07 & & & 24443813.40 & 17.27 &  0.17 \\
 24437382.64 & 15.53 &  0.11 & & & 24443817.48 & 17.99 &  0.06 \\
 24439093.44 & 16.59 &  0.07 & & & 24444172.48 & 16.77 &  0.27 \\
 24439144.39 & 16.63 &  0.15 & & & 24444225.39 & 16.86 &  0.21 \\
 24439150.46 & 16.17 &  0.14 & & & 24444225.53 & 17.16 &  0.07 \\
 24439153.51 & 16.64 &  0.12 & & & 24444308.52 & 16.83 &  0.11 \\
 24442005.35 & 16.64 &  0.25 & & & 24444340.41 & 17.22 &  0.06 \\
 24442014.45 & 16.53 &  0.12 & & & 24444582.50 & 17.48 &  0.18 \\
 24442043.45 & 16.61 &  0.33 & & & 24444603.49 & 17.04 &  0.16 \\
 24442067.34 & 16.49 &  0.13 & & & 24444668.42 & 17.24 &  0.27 \\
 24442339.43 & 16.11 &  0.24 & & & 24444934.52 & 17.11 &  0.08 \\
 24442391.34 & 16.77 &  0.19 & & & 24445026.50 & 14.76 &  0.14 \\
 24442520.53 & 16.25 &  0.31 & & & 24445050.33 & 15.81 &  0.06 \\
 24442835.33 & 17.40 &  0.14 & & & 24445293.52 & 15.80 &  0.18 \\
 24443134.50 & 17.17 &  0.22 & & & 24445645.10 & 17.40 &  0.20 \\
 24443163.41 & 17.36 &  0.28 & & & 24445663.53 & 17.29 &  0.18 \\
 24443436.56 & 16.70 &  0.28 & & & 24446091.43 & 16.21 &  0.08 \\
 24443463.49 & 17.57 &  0.17 & & &             &       &       \\
 24443485.36 & 16.70 &  0.38 & & &             &       &       \\
 24443485.37 & 17.65 &  0.26 & & &             &       &       \\
\enddata

\end{deluxetable}

% figura 1

\clearpage

% figura 2
\clearpage
\begin{figure}
\epsscale{1.0}
\caption{Reference stars in the field of S5 0716+714.
The reference stars are marked with small letters and \s~ is marked with S5.
\label{cartina}
}
\end{figure}

%figura 3  

\clearpage

%figura 4

\clearpage

%figura 5

\clearpage

%figura 6

%figura 7


\begin{thebibliography}{}
\bibitem[Abraham(2000)]{Abr00}
Abraham, Z., 2000, A\&A 355, 915
\bibitem[Bach et al.(2005)]{Bac05}
Bach, U., Krichbaum, T. P., Ros, E., Britzen, S., Tian, W. W., Kraus, A.,
Witzel, A., \& Zensus, J. A. 2005, A\&A 433, 815
\bibitem[Barbieri et al.(2003)]{Bar03}
Barbieri C., et al. 2003, Experimental Astronomy 15, 29
\bibitem[Biermann et al.(1981)]{Bie81}
Biermann P., et al. 1981, ApJ 247, L53
\bibitem[Blandford \& Rees(1978)]{BlaRee78}
Blandford, R. D., \& Rees, M. J. 1978, Phys. Scripta 17, 265
\bibitem[Caproni \& Abraham(2004a)]{CapAbr04a}
Caproni, A., \& Abraham, Z. 2004a, MNRAS 349, 1218 
\bibitem[Caproni \& Abraham(2004b)]{CapAbr04b}
Caproni, A., \& Abraham, Z. 2004b, ApJ 602, 625 
\bibitem[Gabuzda et al.(1998)]{Gab98}
Gabuzda, D. C., Kovalev, Y. Y., Krichbaum, T. P., Alef, W., Kraus, A.,
Witzel, A., \& Quirrenbach, A. 1998, A\&A 333, 445
%\bibitem[Ghisellini et al. (1997)]{Ghi97}
%Ghisellini G., Villata M., Raiteri C. M. et al. 1997, A\&A 327, 61
\bibitem[Giommi et al.(1999)]{Gio99}
Giommi P., et al. 1999, A\&A 351, 59
\bibitem[Heidt \& Wagner(1996)]{HeiWag96}
Heidt, J., \& Wagner, S. J. 1996, A\&A 305, 42
\bibitem[Kadler et al.(2004)]{Kad04}
%Kadler, M., Kerp, J., \&  Krichbaum, T. P. 2004, A\&A submitted
Kadler, M., Ros, E., Kerp, J., Kovalev, Y., \&  Zensus, J. A. 2004,
Proc. 7th EVN Symp., Toledo Oct. 2004,
(R. Bachiller, F. Colomer, J.-F. Desmurs, P. de Vincente eds.), p. 23
\bibitem[Katz(1997)]{Kat97}
Katz, J. I. 1997, ApJ 478, 527
\bibitem[Massaro et al.(1999)]{Mas99}
Massaro, E., et al. 1999, A\&A 342, L49
\bibitem[Massaro et al.(2001)]{Mas01}
Massaro, E., Mantovani, F., Fanti, R., Nesci, R., Tosti, G., \& Venturi, T. 
2001, A\&A 374, 435
\bibitem[Massaro \& Mantovani(2004)]{MasMan04}
Massaro, E., \& Mantovani, F. 2004, Proc. 7th EVN Symp., Toledo Oct. 2004,
(R. Bachiller, F. Colomer, J.-F. Desmurs, P. de Vincente eds.), p. 39
\bibitem[Massaro et al.(2004)]{Mas04}
Massaro, E., et al. 2004, A\&A 423, 935
\bibitem[Miller(1978)]{Mil78}
Miller, H. R. 1978, ApJ 223, L67
\bibitem[Nesci et al.(1997)]{Nes97}
Nesci, R., Aniello, T., Fiorucci, M., Maesano, M., Massaro, E., Montagni, F., 
\& Tosti, G. 1997, Mem SAIt 68, 207
\bibitem[Ostorero \& Wagner(2004)]{Ost04}
Ostorero, L., \& Wagner, S. 2004, 

http://www.lsw.uni-heidelberg.de/users/swagner/Efiles/proceedingsEM4.pdf    
%\bibitem[Raiteri et al. (2001)]{Rai01}
%Raiteri C.M., Villata M., Aller H.D. et al. 2001, A\&A 377, 396
\bibitem[Raiteri et al.(2003)]{Rai03}
Raiteri, C. M., et al. 2003, A\&A 402, 151
\bibitem[Savolainen et al.(2002)]{Sav02}
Savolainen, T., Wiik, K., Valtaoja, E., Jorstad, S. G., \& Marscher, A. P.
2002, A\&A 394, 851
\bibitem[Sillanp\"a\"a et al.(1988)]{Sil88}
Sillanp\"a\"a, A., Haarala, S., Valtonen, M. J., Sundelius, B., \& Byrd, G. G. 
1988, ApJ 325, 628
\bibitem[Stirling et al.(2003)]{Sti03}
Stirling, A. M., et al., 2003 MNRAS 341, 405 
\bibitem[Tateyama \& Kingham(2004)]{TatKin04}
Tateyama, C. E., \& Kingham, K. A. 2004, ApJ 608, 149
\bibitem[Villata et al.(1998)]{Vil98}
Villata, M., Raiteri, C. M., Lanteri, L., Sobrito, G., \& Cavallone, M. 1998, 
A\&AS 130, 305
\bibitem[Villata et al.(2000)]{Vil00}
Villata, M., et al. 2000, A\&A 363, 108 
\bibitem[Wagner et al.(1996)]{Wag96}
Wagner, S. J., et al. 1996, AJ 111, 2187
\end{thebibliography}
\end{document}